\documentclass[pra,twocolumn,amsmath,amssymb]{revtex4}
\usepackage{graphicx}
\usepackage{bm}
\usepackage{amsmath,amssymb}
\usepackage{color}

\begin{document}

\title{Verification for measurement-only blind quantum computing}  

\author{Tomoyuki Morimae}
\affiliation{ASRLD Unit, Gunma University, 1-5-1 Tenjin-cho,
Kiryu-shi, Gunma 376-0052, Japan}
\affiliation{Department of Physics, Imperial College London, London SW7 2AZ, United Kingdom}
\date{\today}
            
\begin{abstract}
Blind quantum computing is a new secure quantum computing protocol
where a client who does not have any sophisticated quantum technlogy
can delegate her quantum computing to a server without leaking any privacy.
It is known that a client who has only a measurement device can perform
blind quantum computing [T. Morimae and K. Fujii, Phys. Rev. A {\bf87}, 050301(R)
(2013)].
It has been an open problem whether the protocol can enjoy the verification,
i.e., the ability of client to check the correctness of the computing.
In this paper, we propose a protocol of verification for
the measurement-only blind quantum computing.
\end{abstract}
\maketitle  

\section{introduction}
Blind quantum computing~\cite{BFK,FK,MABQC,Barz,Vedran,
composableV,
Lorenzo,Joeoptimal,Barz2,NandV} 
is a secure delegated quantum computing, where a client (Alice),
who does not have enough quantum technology, delegates
her quantum computing to a server (Bob), who has a fully-fledged
quantum computer, without leaking any information about
her computation to Bob.
A blind quantum computing protocol for almost classical Alice
was first proposed by Broadbent, Fitzsimons, and Kashefi~\cite{BFK}
by using the measurement-based model due to Raussendorf
and Briegel~\cite{MBQC}.
In their protocol, Alice only needs a device which emits randomly
rotated single-qubit states.
Later it was shown that weak coherent pulses, instead of
single-photon states, 
are sufficient for blind quantum computation~\cite{Vedran}.
Recently, it was shown 
that blind quantum computing can be verifiable~\cite{FK,Barz2,NandV}.
Here, verifiable means that Alice can test Bob's computation~\cite{FK,Barz2,
NandV}. The verifiability is an important requirement,
since Alice cannot recalculate the result of the delegated
computation by herself to check the correctness
(remember that she does not have any quantum computer),
and therefore if there is no verification method,
she might be palmed off with a wrong result
by a fishy company who tries to sell a fake 
quantum computer~\cite{Barz2,NandV}.
The verifiable blind protocol was experimentally demonstrated with a photonic
qubit system~\cite{Barz2,NandV}.

Recently, another type of blind quantum computing protocol was proposed
in Ref.~\cite{MABQC}.
In this protocol, Alice needs only a device that can measure
quantum states.
One advantage of this protocol is that the security is device 
independent~\cite{DI,DI2,BCK,BCK2,McKague}, 
and is based on the no-signaling principle~\cite{Popescu}, 
which is more fundamental than quantum physics.
However, it has been an open problem whether the protocol can enjoy verification.

In this paper, we propose a verification protocol for
the measurement-only blind quantum computing.
We will propose two protocols.
Interestingly, our protocols are based on
the combination of 
two different concepts from different fields: 
the no-signaling principle~\cite{Popescu} from the foundation of physics and
the topological quantum error correcting code~\cite{RHG1,RHG2,Kitaev}
from a practical application in quantum information.
The no-signaling principle means that a shared quantum (or more general) state
cannot be used to transmit information.
It is one of the most central principles in physics, and known to
be more fundamental than quantum physics (i.e., there is a theory
which is more non-local than quantum physics but does not violate
the no-signaling principle~\cite{Popescu}).
The topological quantum error correcting code
is a specific type of the quantum error correcting code
which cleverly uses the topological order of exotic quantum
symmetry-breaking systems 
to globally encode logical states.

\section{Topological measurement-based quantum computation}
\begin{figure}[htbp]
\begin{center}
\includegraphics[width=0.4\textwidth]{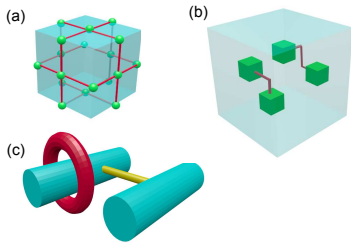}
\end{center}
\caption{
The topological measurement-based quantum computation.
(a) The elementary cell of the Raussendorf-Harrington-Goyal state.
Green balls are qubits, and red bonds are $CZ$ gates.
(b) The error detection. Red strings are errors. Green boxes
are syndrome operators.
(c) Undetected errors or logical operations. Blue tubes are defects. 
Red and yellow strings are strings of operators,
which surround or connect defects, respectively.
} 
\label{DI_fig3}
\end{figure}

The Raussendorf-Harrington-Goyal state $|RHG\rangle$ is the three-dimensional
graph state with the elementary cell given in Fig.~\ref{DI_fig3} (a).
Defects in the graph state are created by $Z$ measurements
on $|RHG\rangle$ as usual in the cluster measurement-based model. 
Topological braidings of defect tubes
can implement some Clifford gates~\cite{RHG1,RHG2,Kitaev}.
Non-Clifford gates, that are necessary for the universal quantum
computation, are implemented by the magic state 
preparation and distillation~\cite{magic}.
A string of $Z$ operators acting on the resource state, which has at least one open edge,
is considered as an error, and its edge(s)
is detected by syndrome measurements
of cubicles of $X$ operators (Fig.~\ref{DI_fig3} (b)).
A string of $Z$ operators on the resource states, which connects
or surrounds defects (Fig.~\ref{DI_fig3} (c)),
is not detected, and can be a logical error.
Local adaptive measurements can implement quantum
computation as well as syndrome error detection.

\section{First protocol}
\begin{figure}[htbp]
\begin{center}
\includegraphics[width=0.3\textwidth]{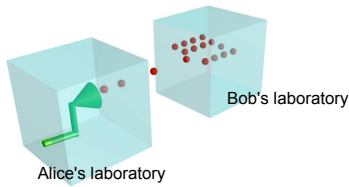}
\end{center}
\caption{
Our setup.
Bob first prepares a resource state.
Bob next sends each particle to Alice one by one.
Alice measures each particle according to her algorithm.
} 
\label{DI_fig1}
\end{figure}

Let us explain our first protocol.
The basic idea of our protocol is illustrated in Fig.~\ref{DI_fig1}:
Bob prepares the resource state, and Alice performs measurements.

\begin{figure}[htbp]
\begin{center}
\includegraphics[width=0.3\textwidth]{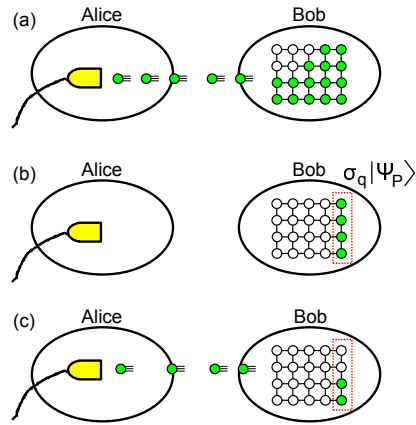}
\end{center}
\caption{
Our protocol. Here, $|\Psi_P\rangle\equiv P(|R\rangle\otimes|+\rangle^{\otimes N/3}
\otimes|0\rangle^{\otimes N/3})$, $P$ is a $N$-qubit permutation, and $|R\rangle$
is a universal resource state.
} 
\label{DI_fig2}
\end{figure}

More precisely, our protocol runs as follows (Fig.~\ref{DI_fig2}).
First, Bob prepares a universal
resource state,
and sends each qubit of it to Alice one by one (Fig.~\ref{DI_fig2} (a)).
Alice measures each qubit until she remotely creates
the $N$-qubit state, $\sigma_q|\Psi_P\rangle$,
in Bob's laboratory (Fig.~\ref{DI_fig2} (b)),
where
$
\sigma_q\equiv \bigotimes_{j=1}^NX_j^{x_j}Z_j^{z_j}
$
with $q\equiv(x_1,...,x_N,z_1,...,z_N)\in\{0,1\}^{2N}$
is the byproduct of the measurement-based quantum computation~\cite{MBQC},
and $X_j$ and $Z_j$ are Pauli operators acting on $j$th qubit.
The state
$
|\Psi_P\rangle\equiv P\Big(|R\rangle\otimes|+\rangle^{\otimes N/3}\otimes|0\rangle^{\otimes N/3}\Big),
$
is the $N$-qubit state,
where $|R\rangle$ is an $N/3$-qubit universal resource state of the
measurement-based quantum computation encoded with a quantum 
error-correcting code of the code distance $d$.
(The size of $|R\rangle$ and the number of traps are optimal, since if
there are too many traps, the efficiency of the computation becomes small,
whereas if there are too few traps, the probability of detecting malicious Bob
becomes small.)
For example, $|R\rangle$ can be the $N/3$-qubit Raussendorf-Harrington-Goyal
state~\cite{RHG1,RHG2} with sufficiently many magic states
being already distilled.
(The Raussendorf-Harrington-Goyal state is the resource state
of the topological measurement-based quantum computing~\cite{RHG1,RHG2}.
In stead of the RHG state, any other quantum error correcting code
can be utilized. Therefore, 
we can also assume that $|R\rangle$ is a normal resource state of
the measurement-based quantum computation encoded with a quantum error-correcting
code.)
We define $|+\rangle\equiv\frac{1}{\sqrt{2}}(|0\rangle+|1\rangle)$,
and $P$ is an $N$-qubit permutation, which keeps
the order of qubits in $|R\rangle$. 
This permutation is randomly chosen by Alice and kept secret to Bob.

Throughout this paper, we assume that there is no communication
channel from Alice to Bob.
Then, due to the no-signaling principle, Bob cannot learn anything
about $P$~\cite{MABQC}. 
If Bob can learn something about $P$,
Alice can transmit some message to Bob by encoding her message
into $P$, which contradicts to
the no-signaling principle. 

Bob sends each qubit of $\sigma_q|\Psi_P\rangle$ to Alice
one by one,
and Alice does the measurement-based quantum computation
on $\sigma_q|\Psi_P\rangle$ with correcting $\sigma_q$ 
(Fig.~\ref{DI_fig2} (c)).
This means that before measuring $j$th qubit of $\sigma_q|\Psi_P\rangle$
she applies $\sigma_q^\dagger|_j$ on $j$th qubit,
where
$\sigma_q^\dagger|_j$ is the restriction
of $\sigma_q^\dagger$ on $j$th qubit.
For example, $(I\otimes XZ\otimes Z)|_2=XZ$.
Qubits belonging to $|R\rangle$ are used to implement
the Alice's desired quantum computation.
States $|0\rangle$ and $|+\rangle$
are used as ``traps"~\cite{FK}.
In other words,
she measures $Z$ on $|0\rangle$ and $X$ on $|+\rangle$,
and if she obtains the minus result (i.e., $|1\rangle$ or $|-\rangle$ state),
she aborts the protocol.
If results are plus for all traps, she accepts the result
of the measurement-based quantum computation on $|R\rangle$.

\section{Verifiability}
Now we show that if all measurements on traps show the correct results,
the probability that a logical state of Alice's computation is changed
is exponentially small.
In other words, the probability that Alice is fooled by Bob is exponentially small.
Hence, our protocol is verifiable.

Since Bob might be dishonest, he might deviate from the above procedure.
His general attack
is a creation of a different state $\rho$
instead of $\sigma_q|\Psi_P\rangle$.
If he is honest, 
$\rho=\sigma_q|\Psi_P\rangle\langle \Psi_P|
\sigma_q^\dagger$.
If he is not honest, $\rho$ can be any state.
However, for any $N$-qubit state $\rho$, 
there exists a completely-positive-trace-preserving (CPTP) map 
which satisfies
$
\rho=\sum_jE_j\sigma_q|\Psi_P\rangle\langle\Psi_P|
\sigma_q^\dagger E_j^\dagger,
$
where 
$E_j\equiv\sum_\alpha C_j^\alpha\sigma_\alpha$,
is a Kraus operator of the CPTP map,
and $C_j^\alpha$ is a complex number (see Appendix~\ref{App:CPTP}).
Since $E_j^\dagger E_j$ is a POVM,
$
I=\sum_jE_j^\dagger E_j
=\sum_j\sum_{\alpha,\beta}C_j^{\alpha *}C_j^\beta\sigma_\alpha^\dagger
\sigma_\beta,
$
we obtain 
$\sum_j\sum_\alpha|C_j^\alpha|^2=1$.

Bob does not know $q$. Therefore, from Bob's view point,
the state is averaged over all $q$:
\begin{eqnarray}
&&\frac{1}{4^N}\sum_q\sum_j 
\sigma_q^\dagger E_j\sigma_q|\Psi_P\rangle\langle\Psi_P|
\sigma_q^\dagger 
E_j^\dagger \sigma_q\nonumber\\
&=&
\frac{1}{4^N}
\sum_q\sum_{j,\alpha,\beta}C_j^\alpha C_j^{\beta *} 
\sigma_q^\dagger \sigma_\alpha \sigma_q|\Psi_P\rangle\langle\Psi_P|
\sigma_q^\dagger \sigma_\beta^\dagger \sigma_q\nonumber\\
&=&
\frac{1}{4^N}
\sum_q\sum_{j,\alpha}|C_j^\alpha|^2  
\sigma_q^\dagger \sigma_\alpha \sigma_q|\Psi_P\rangle\langle\Psi_P|
\sigma_q^\dagger \sigma_\alpha^\dagger \sigma_q\nonumber\\
&=&
\sum_{j,\alpha}|C_j^\alpha|^2 
\sigma_\alpha |\Psi_P\rangle\langle\Psi_P|\sigma_\alpha^\dagger\nonumber\\
&=&
\sum_\alpha\tilde{C}_\alpha
\sigma_\alpha |\Psi_P\rangle\langle \Psi_P|\sigma_\alpha^\dagger,
\label{attack}
\end{eqnarray}
where
$\tilde{C}_\alpha\equiv\sum_j|C_j^\alpha|^2$
and
$
\sum_\alpha\tilde{C}_\alpha=
\sum_\alpha\sum_j|C_j^\alpha|^2
=1.
$
Here, we have used the following equations~\cite{Aharonov}
\begin{eqnarray}
\sum_q \sigma_q^\dagger\sigma_\alpha\sigma_q\rho \sigma_q^\dagger\sigma_\beta^\dagger\sigma_q&=&0
\label{formula}\\
\frac{1}{4^N}\sum_q \sigma_q^\dagger\sigma_\alpha\sigma_q\rho \sigma_q^\dagger\sigma_\alpha^\dagger\sigma_q&=&
\sigma_\alpha\rho \sigma_\alpha^\dagger\nonumber
\end{eqnarray}
for any $\rho$ and $\alpha\neq\beta$.
The second equation is easy to show.
For a proof of Eq.~(\ref{formula}), see Appendix~\ref{App:diagonalize}.
Equation (\ref{attack}) shows that we can assume that Bob's attack is
the ``random Pauli" attack, i.e., Bob randomly applies Pauli operators on each qubit.

Bob's attacks after creating $\rho$ can also
be included in the preparation
of $\rho$.
This is understood as follows.
Let us assume that, after creating $\rho$, 
Bob sends a subsystem $S_1$ of $\rho$
to Alice, and then Alice measures all particles of $S_1$.
After this Alice's measurement, Bob might apply an operation
on another subsystem $S_2$ of $\rho$ which has not been sent to Alice.
However, Bob cannot know Alice's measurement angles and results on $S_1$
due to the no-signaling principle, and therefore Bob's 
operation on $S_2$ is independent of Alice's measurements
on $S_1$. 
Furthermore, Bob's operation on $S_2$ commutes with Alice's measurements
on $S_1$. 
Hence we can consider as if
Bob applied such an operation on $S_2$ immediately after
he preparing $\rho$.

In short, we can assume that 
Bob's attack is a random Pauli attack on the correct
state $|\Psi_P\rangle$ as is shown in Eq.~(\ref{attack}).
Hence 
let us focus on $\sigma_\alpha|\Psi_P\rangle$.
For many quantum
error correcting code (such as the topological one~\cite{RHG1,RHG2}),
if the weight $|\alpha|$ of $\sigma_\alpha$
is less than a certain integer $d$ (the code distance), 
then such an error is detected or does not
change logical states~\cite{RHG1,RHG2,Kitaev,FK}.
For example, in the topological code, $d$ is determined by the defect thickness and
distance between defects~\cite{RHG1,RHG2}.
Here, the weight $|\alpha|$ of $\sigma_\alpha$ means 
the number of nontrivial operators in $\sigma_\alpha$.
(For example, the weight of $I\otimes XZ\otimes Z\otimes I\otimes X$
is 3.)
Therefore, in order for $\sigma_\alpha$ to change a logical state
of the computation,
$|\alpha|$ must be larger than $d$.
(To understand it, let us consider a simple example. If we encode 
the logical 0 as $|0_L\rangle\equiv|000\rangle$
and the logical 1 as $|1_L\rangle\equiv|111\rangle$,
we must flip more than two qubits to change the logical state.
A single bit flip is detected and corrected when the majority vote
is done.)

Alice randomly chooses a permutation $P$.
In this case, the probability of $P^\dagger\sigma_\alpha P$ 
not changing any trap
is at most $(\frac{2}{3})^{|\alpha|/3}$.
(For a calculation, see Appendix~\ref{App:prob}).
Therefore,
the probability that the logical state is changed and
no trap is flipped
is at most
$
\sum_{|\alpha|\ge d}\tilde{C}_\alpha \left(\frac{2}{3}\right)^{|\alpha|/3}
\le
\left(\frac{2}{3}\right)^{d/3}\sum_{|\alpha|\ge d}\tilde{C}_\alpha
\le
\left(\frac{2}{3}\right)^{d/3},
$
where we have used the fact 
$\tilde{C}_\alpha\ge0$
and
$\sum_{|\alpha|\ge d}\tilde{C}_\alpha\le\sum_\alpha \tilde{C}_\alpha=1$.
Here, we have said ``at most", since 
the above sum includes
the contribution from
$\sigma_\alpha$ which has a weight larger than $d$ but
does not contain any logical error.
In this way, we have shown that the probability that Alice is fooled
by Bob is exponentially small ($d$ can be sufficiently large
by concatenating the code).
As we have seen, no communication from Alice to Bob
is required for the verification. Therefore,
whatever Alice's measurement device does, Bob cannot learn Alice's
computational information because of the no-signaling principle.
In other words,
the security of the protocol is device-independent.


\section{Second protocol}
Let us explain our second protocol, which uses the property of the topological code,
and does not use
any trap.
Alice randomly chooses 
$k\equiv(h_1,...,h_N,t_1,...,t_N)\in\{0,1\}^{2N}$,
and defines the $N$-qubit operator 
$
K_k\equiv \bigotimes_{j=1}^NT_j^{t_j}H_j^{h_j},
$
where $T\equiv |0\rangle\langle0|+i|1\rangle\langle1|$ 
and $H$ is the Hadamard operator.
Note that $T^\dagger XT=-iXZ$, $T^\dagger ZT=Z$, and $T^\dagger XZ T=-iX$.
Next, Alice defines the $N$-qubit state
$|\Psi_k\rangle\equiv K_k|RHG'\rangle$,
where $|RHG'\rangle$ is the $N$-qubit Raussendorf-Harrington-Goyal
state~\cite{RHG1,RHG2}
with
sufficient number of magic states being already distilled~\cite{RHG1,RHG2}.

Bob prepares a universal resource state, and sends each qubit of it to Alice
one by one.
Alice does measurements and creates
$\sigma_q|\Psi_k\rangle$ in Bob's laboratory,
where 
$\sigma_q$ is the byproduct of the measurement-based quantum
computation.
Due to the no-signaling principle, Bob cannot learn $k$.
Bob sends each qubit of $\sigma_q|\Psi_k\rangle$
to Alice one by one, and Alice does her topological measurement-based
quantum computation with correcting $\sigma_qK_k$.
If Alice detects any error, she aborts the protocol.

Again, because of Eq.~(\ref{attack}),
we can assume that Bob's attack is a random Pauli attack.
Therefore let us focus on $\sigma_\alpha|\Psi_k\rangle$.
In order for $\sigma_\alpha$ to change a logical state
without being detected by syndrome measurements,
$\sigma_\alpha$ must contain at least one
string $s_\alpha$ of operators 
which connects or surrounds defects 
(Fig.~\ref{DI_fig3} (c))~\cite{RHG1,RHG2,Kitaev}.  
Since Alice randomly chooses $k$, the probability that 
all operators in $K_k^\dagger s_\alpha K_k$
become $Z$ or $XZ$ operators is at most $(\frac{3}{4})^{|s_\alpha|}$,
where $|s_\alpha|$ is the weight of $s_\alpha$.
Note that $|s_\alpha|\ge d$ because it connects or
surrounds defects.

Hence,
the probability that the logical state is changed
and Alice does not detect any error
is at most
$
\sum_{|\alpha|\ge d}\tilde{C}_\alpha \left(\frac{3}{4}\right)^{|s_\alpha|}
\le
\left(\frac{3}{4}\right)^d\sum_{|\alpha|\ge d}\tilde{C}_\alpha
\le
\left(\frac{3}{4}\right)^d.
$
In short, our second protocol is also verifiable.
Again, the device-independent security is guaranteed
by the no-signaling principle.

\acknowledgements
The author acknowledges 
supports by JSPS and Tenure Track System by MEXT Japan.

\appendix

\section{Existence of a CPTP map}
\label{App:CPTP}
Let  
$\{|\phi_k\rangle\}_{k=1}^{2^N}$
be any orthonormal basis
of the $N$-qubit Hilbert space,
$\langle\phi_k|\phi_j\rangle=\delta_{k,j}$.
We diagonalize the $N$-qubit state $\rho$ as 
$\rho=\sum_{j=1}^{2^N}\lambda_j|\lambda_j\rangle\langle\lambda_j|$.
Let us take
$E_{jk}\equiv\sqrt{\lambda_j}|\lambda_j\rangle\langle\phi_k|$.
Then for any $N$-qubit state $\eta=\sum_{\alpha,\beta}\eta_{\alpha\beta}
|\phi_\alpha\rangle\langle\phi_\beta|$,
$
\sum_{j,k}E_{jk}\eta E_{jk}^\dagger
=
\sum_{j,k,\alpha,\beta}\sqrt{\lambda_j}\sqrt{\lambda_j}
\eta_{\alpha\beta}
|\lambda_j\rangle\langle\phi_k|\phi_\alpha\rangle
\langle\phi_\beta|\phi_k\rangle\langle\lambda_j|
=
\sum_{j,k}\lambda_j
\eta_{kk}
|\lambda_j\rangle
\langle\lambda_j|
=
\rho.
$
Furthermore,
$
\sum_{j,k}E_{jk}^\dagger E_{jk}
=
\sum_{j,k}\sqrt{\lambda_j}\sqrt{\lambda_j}
|\phi_k\rangle\langle\lambda_j|
\lambda_j\rangle\langle\phi_k|
=
\sum_{j,k}\lambda_j
|\phi_k\rangle\langle\phi_k|
= I.
$

\section{Proof of Eq.~(\ref{formula})}
\label{App:diagonalize}
For the convenience of readers, we here give the proof~\cite{Aharonov}
of Eq.~(\ref{formula}).
Since $\alpha\neq\beta$, there exists an index $j$ such that 
$\sigma_\alpha|_j\neq \sigma_\beta|_j$.
For any such
$\sigma_\alpha|_j$ and $\sigma_\beta|_j$,
we can always 
take $S\in\{X,Z\}$ such that $S$ anticommutes only one of 
$\sigma_\alpha|_j$ and $\sigma_\beta^\dagger|_j$.
Let us define 
$Q\equiv I^{\otimes j-1}\otimes S \otimes I^{\otimes N-j}$.
Then,
$
\sum_q \sigma_q^\dagger\sigma_\alpha\sigma_q\rho\sigma_q^\dagger \sigma_\beta^\dagger\sigma_q
=
\sum_q (Q\sigma_q)^\dagger\sigma_\alpha(Q\sigma_q)
\rho(Q\sigma_q)^\dagger \sigma_\beta^\dagger(Q\sigma_q)
=
\sum_q (\sigma_q^\dagger Q^\dagger)\sigma_\alpha(Q\sigma_q)
\rho(\sigma_q^\dagger Q^\dagger)\sigma_\beta^\dagger(Q\sigma_q)
=
-\sum_q \sigma_q^\dagger \sigma_\alpha \sigma_q
\rho \sigma_q^\dagger \sigma_\beta^\dagger \sigma_q.
$

\section{Probability of avoiding traps}
\label{App:prob}
Let $\sigma_{\alpha}|_j$ be the restriction of $\sigma_\alpha$
for $j$th qubit. (For example, $(X\otimes I\otimes XZ)|_3$ is $XZ$.)
Let $a$, $b$, $c$ be the number of $j$ such that
$\sigma_\alpha|_j=X$, $Z$, $XZ$, respectively.
(In other words, $a$ is the number of $X$ operators in $\sigma_\alpha$,
$b$ is the number of $Z$ operators in $\sigma_\alpha$,
and $c$ is the number of $XZ$ operators in $\sigma_\alpha$.)
We define $|\alpha|$ be the weight of $\sigma_\alpha$, namely,
the number of non-$I$ operators.
Since
$|\alpha|=a+b+c\le3\max(a,b,c)$,
we obtain
$\max(a,b,c)\ge\frac{|\alpha|}{3}$.

Let us assume $\max(a,b,c)=a$.
Then, the probability that all $X$ operators of $\sigma_\alpha$
do not change any trap is
$
\frac{(N-a)!\prod_{k=0}^{a-1}(\frac{2N}{3}-k)}{N!}
=
\left(\frac{2}{3}\right)^a
\frac{\prod_{k=0}^{a-1}(N-\frac{3}{2}k)}{\prod_{k=0}^{a-1}(N-k)}
\le
\left(\frac{2}{3}\right)^a
\le
\left(\frac{2}{3}\right)^{|\alpha|/3}.
$
This is larger than the probability that
$\sigma_\alpha$ does not change any trap.

We can obtain the same result for $\max(a,b,c)=b$. 
For $\max(a,b,c)=c$, we have only to replace $\frac{2}{3}$
with $\frac{1}{3}$.

\end{document}